\documentclass[sigplan,10pt,balance]{acmart}
\usepackage{booktabs}   
\usepackage{subcaption} 
\usepackage{afterpage}
\usepackage{booktabs} 
\usepackage{setspace}
\usepackage{url}
\usepackage{alltt}
\usepackage{xspace}
\usepackage{epsfig}
\usepackage{graphics}
\usepackage{bbold}
\usepackage{empheq}
\usepackage{caption}
\usepackage{colortbl,hhline}
\usepackage{multirow}
\usepackage{tikz}
\usepackage{tikz-qtree}
\usepackage{braket}
\usepackage{comment}
\usepackage{pdfpages}

\usepackage{amsmath,amsfonts,amsbsy}
\usepackage{fancyvrb,keyval,ifthen}
\usepackage{graphicx}
\usepackage{qcircuit}

\usepackage{listings}
\usepackage{enumitem}
\setlist{nolistsep}
\usepackage{balance}
\usepackage{array}

\usepackage[utf8]{inputenc}

\usepackage{algorithm}
\usepackage{algorithmic}

\definecolor{mGreen}{rgb}{0,0.6,0}
\definecolor{mGray}{rgb}{0.5,0.5,0.5}
\definecolor{mPurple}{rgb}{0.58,0,0.82}
\definecolor{backgroundColour}{rgb}{0.95,0.95,0.92}

\lstdefinestyle{cstyle}{
    commentstyle=\color{mGreen},
    keywordstyle=\color{magenta},
    numberstyle=\tiny\color{mGray},
    stringstyle=\color{mPurple},
    breakatwhitespace=false,
    breaklines=true,
    escapeinside={(*@}{@*)},
    captionpos=b,
    keepspaces=true,
    numbers=left,
    numbersep=5pt,
    showspaces=false,
    showstringspaces=false,
    showtabs=false,
    tabsize=1,
    language=C
}

\newcommand{\mytitle}[0]{String Abstractions for Qubit Mapping}
\newcommand{\tket}[0]{\texttt{TKET}}
\newcommand{\qiskit}[0]{\texttt{QISKIT}}
\newcommand{\pyquil}[0]{\texttt{Pyquil}}
\newcommand{\cirq}[0]{\texttt{Cirq}}
\newcommand{\sbqa}[0]{\texttt{SBQA}}
\newcommand{\muqut}[0]{\texttt{MUQUT}}

\newcommand{\mss}[0]{\texttt{SS}}
\newcommand{\mgf}[0]{\texttt{GF}}

\newcommand{\mssb}[0]{\texttt{SS}$_{b}$}
\newcommand{\mgfb}[0]{\texttt{GF}$_{b}$}
\newcommand{\mgsfb}[0]{\texttt{GSF}$_{b}$}
\newcommand{\msso}[0]{\texttt{SS}$_{o}$}
\newcommand{\mgfo}[0]{\texttt{GF}$_{o}$}
\newcommand{\mgsfo}[0]{\texttt{GSF}$_{o}$}

\AtBeginDocument{%
  \providecommand\BibTeX{{%
    \normalfont B\kern-0.5em{\scshape i\kern-0.25em b}\kern-0.8em\TeX}}}

\acmYear{2021}\copyrightyear{2021}
\setcopyright{rightsretained}
\acmConference{}{}{}
\acmBooktitle{}
\acmPrice{}

\begin{document}

\title{\mytitle{}}


\author{Blake Gerard}
\affiliation{%
  \institution{The University of Oklahoma}
  \city{}
  \state{}
  \country{}
  \postcode{}
}
\email{blake.gerard@ou.edu}
\author{Martin Kong}
\affiliation{%
  \institution{The University of Oklahoma}
  \city{}
  \state{}
  \country{}
  \postcode{}
}
\email{mkong@ou.edu}


\renewcommand{\shortauthors}{Gerard and Kong}

\begin{abstract}

One of the key compilation steps in Quantum Computing (QC) is to determine an
initial logical to physical mapping of the qubits used in a quantum circuit.
The impact of the starting qubit layout can vastly affect later scheduling and
placement decisions of QASM operations, yielding higher values on critical
performance metrics (gate count and circuit depth) as a result of quantum
compilers introducing SWAP operations to meet the underlying physical
neighboring and connectivity constraints of the quantum device. 

In this paper we introduce a novel qubit mapping approach, string-based
qubit mapping. The key insight is to prioritize the mapping of logical qubits that
appear in longest repeating non-overlapping substrings of qubit pairs accessed.
This mapping method is complemented by allocating qubits according to their
global frequency usage. We evaluate and compare our new mapping scheme against 
two quantum compilers (QISKIT and TKET) and  two device topologies,
the IBM Manhattan (65 qubits) and the IBM Kolkata (27 qubits). Our results
demonstrate that combining both mapping mechanisms often achieve better results
than either one individually, allowing us to best QISKIT and TKET baselines,
yielding between 13\% and 17\% average improvement in several group sizes, 
up to 32\% circuit depth reduction and 63\% gate volume improvement.

\end{abstract}


\maketitle

\section{Introduction}
\label{sec:intro}

The field of Quantum Computing (QC) is rapidly growing.  Several technologies
and applications are being sought and designed to demonstrate Quantum Supremacy
\cite{quantum.supremacy.nature.2019}.  At the same time, various quantum
computers have been constructed and made accessible to the public as cloud
services \cite{ibm-services}.  Numerous real quantum devices such as Google
Sycamore and Rigetti's Aspen-9 \cite{aspen9.rigetti}  have been proposed to
tackle problems beyond the computational power of today's supercomputers.  This
effort has been complemented by multiple applications being expressed using
frameworks such as OpenFermion \cite{openfermion.arxiv.2017}, while various
quantum compilers (\qiskit{} \cite{qiskit}, Rigetti's \pyquil{}
\cite{pyquil.github}, Google's \cirq{} \cite{cirq.zenodo}, Cambridge's \tket{}
\cite{tket}) support the task of translating and mapping quantum programs onto
real hardware. 

Ultimately, programs are lowered to a form of quantum assembly such as OpenQASM
\cite{openqasm.arxiv.2017} and used as input to the rest of the quantum compilation
process (or ``transpilation'' in \qiskit{} terminology).
Compiling quantum programs involves several steps,
including determining the initial qubit layout, introducing special operations
to move the state of a qubit (Quantum Bit) between non-physically adjacent qubits
(also known as layout synthesis or routing), gate scheduling, and further lowering
the program to technology dependent, native operations available in the physical device,
e.g. pulse compilation \cite{qiskit-pulse.iop.2020} \cite{olsq}. 

In this paper we focus on the problem of determining the initial qubit layout,
that is, the task of assigning each logical qubit to a physical qubit on an 
specific target device. In particular, a quantum computer has a fixed number
of physical qubits, $Q$, which are physically connected by the underlying 
technology. Thus, the quantum device is often represented as a coupling graph
$G=(Q,E)$. The existence of an edge in the graph denotes that the qubits connected
by it can be used as operands to a two-qubit quantum gate operation. In this scenario
we say that the qubits are ``physically adjacent''. However, when this is not the case,
and owing to the ``no cloning theorem'' \cite{no.cloning.theorem},
the compiler is required to introduce special quantum operations, SWAPs, to exchange
the state between two non-physically adjacent qubits. 

Given the number of physical qubits of a device, $Q$, the number of possible
layouts is $Q!$. Given this, the problem of determining a logical-to-physical
qubit mapping
consists on computing a permutation of
the original logical qubit sequence $q_0, q_1, \ldots, q_{n-1}$ so that quality
and performance metrics (e.g. circuit depth, gate volume, total error and
even execution time) are optimized.


We propose a new qubit mapping class, String-Based Qubit Allocation (\sbqa{}).
The key insight is to prioritize qubit allocation over sub-string sequences,
and within these, further favor qubits that are heavily used. Strings are extracted
from a linearized sub-trace of the quantum program consisting of only indices
used in two-qubit gates. Once the {\em high-impact qubits} are processed within
a sub-string, repeated occurrences are removed from the sub-trace before locating
the next repeated sub-string to target. This approach is then complemented with a simple
qubit allocation using global histograms. Our results show that 
our qubit mapping strategies often achieve 13\%-15\% average improvement
on several groups of quantum circuits (arranged by size), peaking at 32\%
circuit depth improvement and 63\% in gate volume.

To summarize, the contributions of our work are:
i) We propose two new qubit mapping techniques, Sub-String (SS) and Global
Frequencies (GF), which are designed to be used both in combination or in stand-alone
fashion. In particular, SS is, to the best of our knowledge, the first qubit
mapping technique driven by string patterns. 
ii) We conduct a comprehensive evaluation of our techniques,
comparing to layout methods available in state-of-the-art quantum
compilers (\qiskit{} and \tket{}). Furthermore, we demonstrate that a two-phase
qubit mapping strategy offers benefits above any of the individual techniques
we propose.
iii) To address the urgency for more scalable techniques, we strongly motivate the
need for novel qubit mapping strategies by identifying limitations of current
qubit/gate mappers. We thus depart from graph- and
solver-based approaches. Moreover, we demonstrate that techniques availabe
today are better tuned for small circuits and small quantum devices.
iv) We introduce a new standalone tool that can be used by the community to explore
new qubit mapping strategies.


The rest of this paper is organized as follows. Section \ref{sec:motiv}
further motivates our research problem. Section \ref{sec:background} covers the
main background and related concepts needed. Section \ref{sec:overview} provides
an overview and working example applying our technique. Section \ref{sec:mapping}
introduces our string-based qubit mapping methods. Section \ref{sec:evaluation}
presents a comprehensive evaluation and comparison of our method against baseline
compilers, initial layout and routing methods. Section \ref{sec:related} recaps
and summarizes the most recent and relevant related work to the problem of qubit
mapping. Finally, Section \ref{sec:conclusion} presents the conclusions and future
directions of our work.

\section{Motivation}
\label{sec:motiv}

Numerous techniques proposed in prior work have simultaneously
addressed the qubit mapping and routing problem. However, we make the distinction
between finding a good initial logical-to-physical qubit mapping (which can
change during the compilation of the program) and the process of introducing
SWAP operations to amend physical connectivity constraints not met naturally
in the program. We thus defer the {\em routing problem} to already
available passes in today's quantum compilers.

To the best of our knowledge, all quantum compilation passes operate over the
dependence graph extracted from the quantum assembly representation. 
\tket{}'s \texttt{Graph Placement} \cite{tket-placement.dagstuhl.2019} attempts
to map logical qubits in a greedy fashion by building graphs from independent
pairs of qubits accessed through various {\em time steps} (a logical partition
of gates that can be executed in parallel). Once the graph is built, it finds
an isomorphic sub-graph in the coupling graph representing the physical device.
In particular, this method ensures that {\em only} two timesteps are free of
SWAP operations. The algorithm also breaks {\em long lines} that fail to map
into smaller ones until they are successfully assigned.  In a similar spirit,
\muqut{} \cite{muqut} attempts to map qubits by extracting sub-graphs of fixed
cardinality from the program's dependence graph and proceeding to map them to
isomorphic sub-graph of the coupling graph. Seed nodes used for extracting the
logical sub-graph are randomly selected.
Maslov, Falconer and Mosca \cite{maslov.tcad.2008} proposed one of the earliest
qubit placement heuristics based on recursive graph partitioning.
Techniques  like \texttt{Dense Layout} \cite{qiskit} in \qiskit{} map qubits by
searching for the ``most connected subset of (logical) qubits''.
Zulehner, Paler and Wille \cite{jku.date.2018} 
partition the input program dependence graph into network layers, and 
apply A* search techniques to find local optimal mappings; Mappings between
adjacent layers are then repaired in a later pass.
Siraichi et al. \cite{siraichi.cgo.2018} select the initial mapping by
determining heavily used logical qubits, and mapping them to physical ones
that most closely match their out-degree. A second class of approaches, which
incur an even higher computational complexity, resort to Integer Linear Programming
and Satisfiability Modulo Theory (SMT) solvers to find optimal solutions;
The qubit (and gates) mapping problem is formulated
to assign individual
gates onto the graph defining the connectivity of a quantum device
\cite{murali.asplos.2019,olsq,wille.dac.2019,murali.asplos.2020}, making it even
harder to decouple a strong inital mapping from gate scheduling and mapping.

\newcommand{\motivcirc}[0]{\texttt{rd84\_253}}
\paragraph{A simple example}
To further motivate the need for better qubit mapping techniques,
consider the \motivcirc{} quantum circuit from the IBM-QX circuit set \cite{jku-ibmqx-web}.
This circuit consists of 13,659 quantum gates, of which 5,960 are two-qubit
gates. Previous studies have shown that two-qubit gates, significantly
impact several quality metrics of quantum programs \cite{olsq,queko} 
owing to the topological constraints
that must be met for a two-qubit gate to be executed. More precisely, 
a two-qubit gate can only be executed on two qubits that are physically
linked in the coupling graph that defines the quantum device. 

We have designed two complementary qubit mapping techniques, one based on {\em
Sub-String} (SS) detection of two-qubit quantum operations, and a {\em Global
Frequency} (GF) approach.  Furthermore, when applied back-to-back, the combined
mapping strategy, {\em
Global Sub-string Frequency} (GSF), often delivers stronger results than any one
individual technique. To demonstrate this, we present in Table \ref{tab:motiv}
excerpts of our
results: the variation in circuit depth (number of quantum operations in the 
critical path), variation in gate volume (number of gates) and variation
in SWAP operations (state exchange between two any qubits). Results are normalized
to the baseline compiler \qiskit{} using the \texttt{SABRE} initial layout method
and \tket{} with {\em Noise-Aware Placement}. For circuit depth and gate volume,
any value above 1.00 represents an improvement, while for the SWAP Ops, positive
values indicate improvement.
Each of our three mapping methods are compiled with a {\em base} configuration
(suffix $_{b}$) and with an {\em optimized} one (suffix $_{o}$) -- Flags used
shown in Section \ref{sec:evaluation}. As we can see, our methods achieve up
to $1.19\times$ circuit depth improvement with \qiskit{} and up to $1.20\times$
with \tket{}. At the same time, with \mgsfo{} on \tket{}, we produce up to
$1.25\times$ improvement on the gate volume metric, out matching all other
mappings.  Lastly, our mapping methods also improves the number of needed
SWAP operations, obtaining a 29\% reduction with \mgsfo{}. 
As a point of reference the best improvement in gate volume reported 
in \cite{sabre} was 30\% on a circuit with 3,888 QOPS (quantum operations)
while for \motivcirc{} the improvement on gate volume was 5\%.
Lastly, the same study was limited to circuits with up to 34,881 gates.

\begin{table}[!htb]
\caption{\label{tab:motiv}\motivcirc{} compilation metrics collected on IBM Manhattan (65 qubit): Circuit Depth, Gate Volume and Inserted SWAP Operations.}
\begin{footnotesize}
\begin{tabular}{l|l|c|c|c|c|c|c}
\toprule
       & Compiler   &  \multicolumn{6}{c}{\bf Mapping Method} \\
Metric & Config.    & SS$_{b}$ & SS$_{o}$ & GF$_{b}$ & GF$_{o}$ & GSF$_{b}$ & GSF$_{o}$\\
\toprule
Circuit & QISKIT & 1.19 & 1.01 & 1.19 & 1.03 & 1.17 & 1.05\\
 Depth & TKET & 1.02 & 1.07 & 0.96 & 1.12 & 1.09 & 1.2\\
\midrule
Gate & QISKIT & 0.67 & 1.01 & 0.7 & 1.04 & 0.65 & 1.07\\
 Volume & TKET & 0.097 & 1.1 & 0.96 & 1.14 & 1.03 & 1.25\\
\midrule
SWAPS & QISKIT & -0.47 & 0.01 & -0.4 & 0.04 & -0.51 & 0.06\\
 Ops & TKET & 0.19 & 0.16 & 0.17 & 0.17 & 0.28 & 0.29\\
\bottomrule
\end{tabular}
\end{footnotesize}
\end{table}

\section{Background}
\label{sec:background}

\paragraph{Qubits}

Qubits are the basic unit of information in quantum computing.
Each qubit represents a linear combination $c_1 |0\rangle + c_2 |1\rangle$ 
of two basis states $|0\rangle$ and $|1\rangle$, with $c_1$ and $c_2$
being complex numbers. A set of logically contiguous qubits conforms
a {\em quantum register}.

\paragraph{Quantum Gates}
Are unitary matrix operations applied to individual qubits.
Current quantum gate typically operate on a single qubit
and on two qubits at the time. For example, the \texttt{CX} (Controlled-NOT)
gate takes two operands, the {\em control qubit} and the {\em target qubit},
while the \texttt{X} gate takes a single qubit.
Each quantum architecture defines a {\em gate basis},
a set of gates to which all programs must be reduced to.
For example, IBM quantum computers use gates \texttt{CX}, \texttt{ID},
\texttt{RZ}, \texttt{SX} and \texttt{X} as their gate basis.

\paragraph{Quantum Assembly and Circuit Diagrams}
Quantum programs can be represented in a device agnostic fashion
using quantum assembly formats such as OpenQASM \cite{openqasm.arxiv.2017}.
Listing \ref{lst:graycode} show the implementation of the \texttt{graycode6\_47}
circuit using OpenQASM. The same program can also be visualized
as a {\em circuit diagram}, as shown in Figure \ref{fig:graycode-circuit}.
Circuit diagrams depict logical qubits as horizontal lanes,
with the time-dimension advancing from left to right.
Single-qubit gates are directly placed on a single lane, while
two-qubit gates connect the qubit lanes used as operands.

\noindent
\begin{minipage}{\linewidth}
\begin{minipage}{0.4\linewidth}
\begin{lstlisting}[basicstyle=\footnotesize]
OPENQASM 2.0;
qreg q[16];
creg c[16];
cx q[1],q[0];
cx q[2],q[1];
cx q[3],q[2];
cx q[4],q[3];
cx q[5],q[4];
\end{lstlisting}
\captionof{lstlisting}{\label{lst:graycode}Graycode in OpenQASM.}
\end{minipage}
\hfill
\begin{minipage}{0.4\linewidth}
\includegraphics[width=\linewidth]{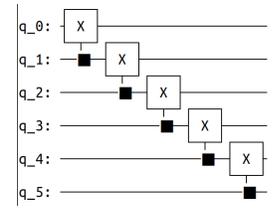}
\captionof{figure}{\label{fig:graycode-circuit}Graycode6\_47 circuit diagram.}
\end{minipage}
\end{minipage}

\paragraph{Coupling Graph}

\begin{figure}[h]
\includegraphics[width=0.45\linewidth]{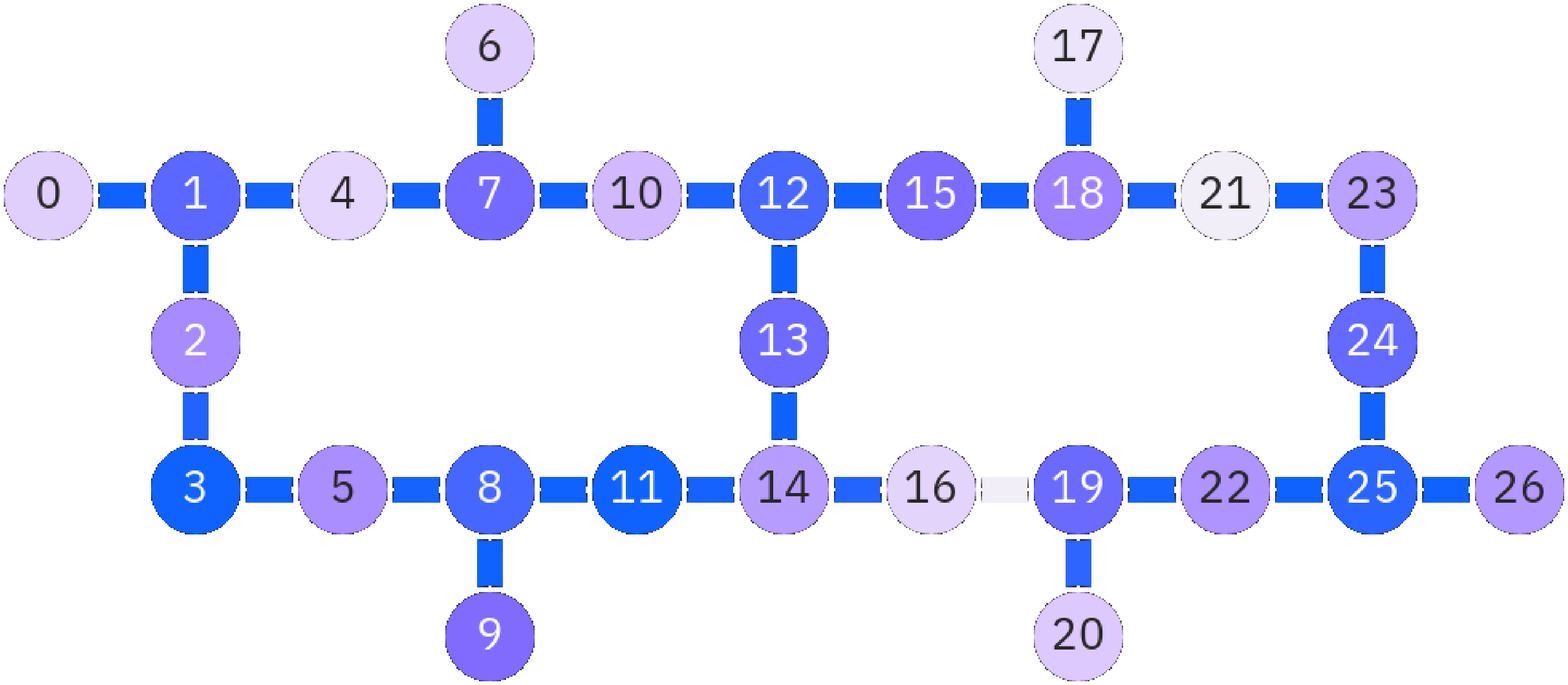}
\includegraphics[width=0.45\linewidth]{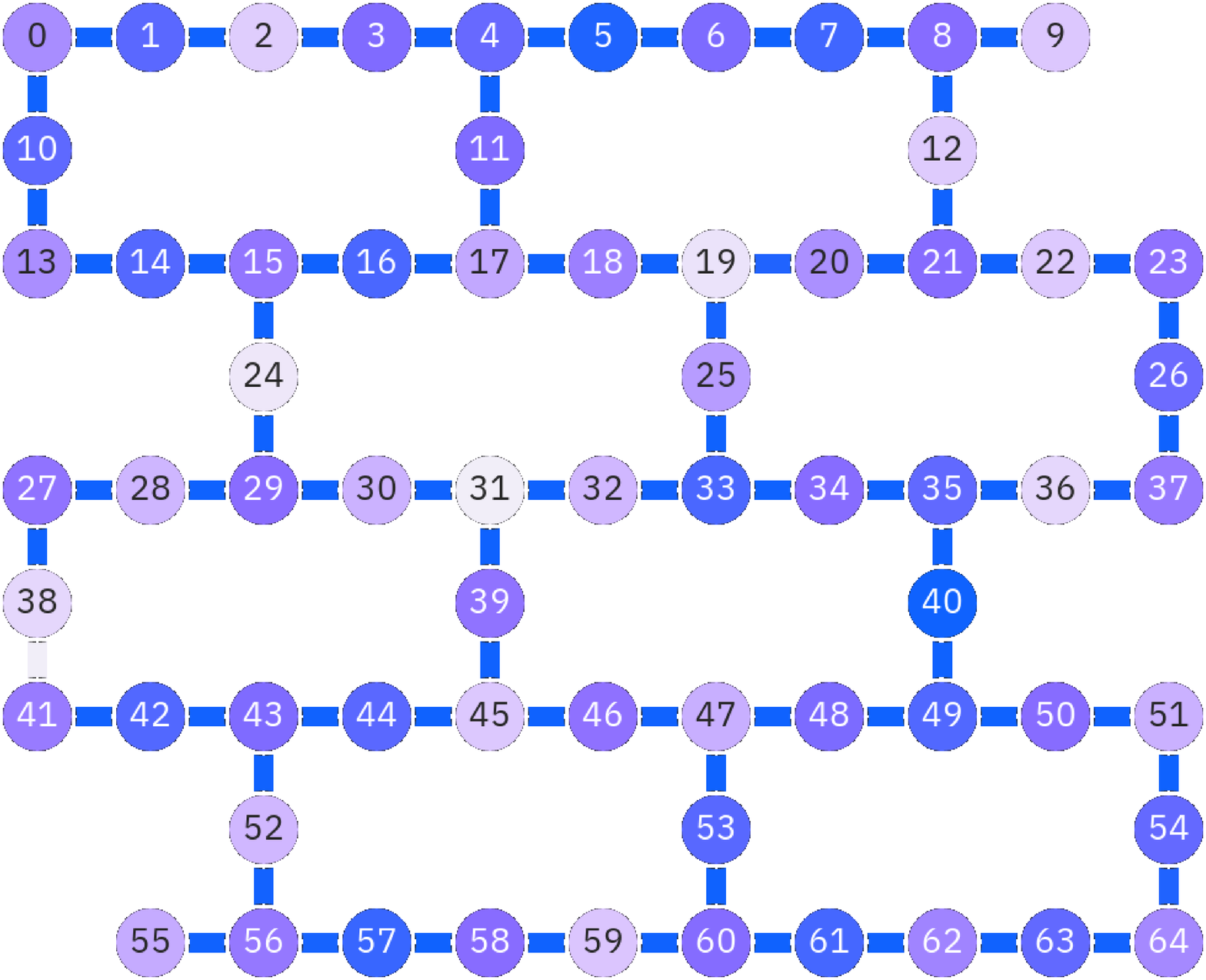}
\caption{\label{fig:coupling-maps}Example coupling maps: 27-qubit device IBM Kolkata (left) and 65-qubit IBM Manhattan (right)}
\end{figure}

Current quantum computers present stringent architectural constraints
that require qubits used as arguments to two-qubit gates (e.g. the \texttt{cx} 
gate) to be qubits that are physically connected. Thus, quantum devices are
conceptually modeled in {\em coupling graphs}. A {\em physical} qubit
in modeled in the graph as a node, while edges linking two nodes establish
the physical connection. Figure \ref{fig:coupling-maps} show the topology
of the quantum devices used in our evaluation, the 27-qubit
IBM Kolkata and the 65-qubit IBM Manhattan \cite{ibm-services}.

\section{Overview}
\label{sec:overview}

\newcommand{\cmax}[0]{\texttt{max46\_240}}
In this section we give a brief overview of the major
steps followed for computing a more effective initial logical-to-physical
qubit mapping. The input to our mapping method is a QASM file
containing the quantum operations (QOPS), such as the one shown
in Fig.~\ref{lst:graycode}. Next, we extract a sub-trace containing
not only the CX (CNOT) gates, but all two-qubit gate operations.
Take for instance circuit \cmax{} from the IBM-QX quantum circuit set
\cite{jku-ibmqx-web}, which consists of 27,126 QOPS,
of which 11,844 are CX gates. The remaining gates are
distributed as 5,076 TDG gates, 6,768 T gates, 3,384 H
gates and 54 X gates. 
Next, for each 2-qubit gate of the form $OPCODE~q_1,~q_2$
we compute a linearized identifier $id_{lin} =q_1~.~ DEV_{width} + q_2$,
where $DEV_{width}$ is the number of physical qubits available in target
quantum device. Thus, all symbols conforming our strings will be defined
in the alphabet $\{ u~.~ DEV_{width} +  v : 0 \leq u,v < DEV_{width} \}$.

Once the program has been converted into a linearized string representation,
we compute a few statistics 
to drive the qubit mapping process.
For example, this program references 76 distinct characters of our alphabet
from the potential 729 ($27^2$). These are all the logical qubit-to-qubit
connections used the circuit. In total, the 76 logical links are used 11,844
times; The average logical connection is used 155.8 times, with a standard
deviation of 174.6. However, not all qubits are used in the same fashion
nor at the same frequency. Consider Table \ref{tab:overview:histogram}, where
for each logical qubit used in circuit \cmax{} we show in the third row, the number of times
that the said qubit is used as the first operand (the control qubit) in a
2-qubit gate. The second row states the number of distinct logical connections
used in the program for each logical qubit, while the fourth and last row
limit the total number of references to only the two and three most frequently
qubits paired with a given qubit $q_i$. From here on, we will often
refer to all pairs $\langle q_i, q_j \rangle$ of a given qubit $q_i$
as the logical pair, edge, link or connections of $q_i$.


\begin{table}[h]
\begin{footnotesize}
\caption{\label{tab:overview:histogram} Edge histogram of circuit \cmax{}.}
\begin{tabular}{c|r|r|r|r|r|r|r|r}
\toprule
Qubit & 
0 & 
1 & 
2 & 
3 & 
6 & 
7 & 
8 & 
9  
\\
\midrule
No. Distinct & 
2 & 
6 & 
8 & 
8 & 
9 & 
9 & 
9 & 
9  
\\
Used Total & 
420  & 
1222 & 
1128 & 
972  & 
1010 & 
1636 & 
2060 & 
1876 
\\
Top 2 &
420  & 
1074 & 
732  & 
460  & 
382  & 
752  & 
1304 & 
1188 
\\
Top 3 &
420  &
1142 &
912  &
580  &
558  &
972  &
1644 &
1360 \\
\bottomrule
\end{tabular}
\end{footnotesize}
\end{table}

Next, we proceed to determine if there
exists a substring pattern in the linearized string that repeats and which
does not overlap with following occurrences. For example, from the linearized
string representation of \cmax{} a substring of length 364 meeting these conditions 
is found 2 times at locations 4,536 and 11,200. This means that there are two sequences
of 364 2-qubit gate operations which can be optimized (hopefully) by the quantum compiler
only once. The repeated sequence found, from a graph-based mapping approach, could
eventually be found as a path (or {\em line} in \tket{} terminology) in the data dependence
graph, but interspersed with single-qubit gates, making the search for gate sequences 
of this length much harder to find.

The actual logical to physical mapping of qubits then proceeds. 
We compute the information listed in Table \ref{tab:overview:histogram} but restricted
to the logical qubits used within the subsequence. Mapping within the gate subsequence
will be constrained to only the qubits whose frequency of usage is above the mean 
value of the subsequence. For example, when limiting this information to only the 
qubit pairs referenced in the substring, the mean value becomes 849.2; A quick
inspection reveals that 5 logical qubits have more logical links than the mean,
namely $q_1, q_2, q_7, q_8$ and $q_9$.
Qubits that satisfy this condition belong to the {\em High-Impact Qubits} (HIQ) set.
Furthermore, the steps that follow will limit the mapping of logical qubits that belong to the
HIQ set. This is done to avoid mapping qubits that will ultimately have a low impact
on later compilation metrics, such as the {\em gate volume}, {\em circuit depth} and
{\em swap count}. Mapping takes place by choosing a {\em leading logical qubit}
and a {\em leading physical qubit} to which to map it. By extension, logical qubits
used as the target node to a logical link will be mapped to the qubits physically adjacent
to the {\em lead physical qubit}. Mapping proceeds by sorting and aligning the
neighboring qubits in both cases. In our example, logical qubit $q_8$ and physical
qubit $q_{25}$ would be chosen as the lead qubits, while logical qubits $q_1$,
$q_7$ and $q_9$ would be mapped to physical qubits $q_{22}$, $q_{24}$ and $q_{26}$.
We note here that logical qubit $q_9$ was also in the HIQ set, and was a close second choice of 
a {\em lead qubit}. However, selecting $q_8$ will preclude choosing $q_9$ later.

Our algorithm will attempt to continue mapping qubits from the HIQ set if the
chosen {\em lead logical qubit} has at least one non-mapped neighbor. If no further
mappings take place within the current string subsequence, the current sequence
is removed from the linearized string representation prior to repeating the
steps starting with the local histogram computation. If no repeating subsequence is found,
the string-based qubit allocation strategy terminates, and we proceed to use only
the global pair frequencies (See Table \ref{tab:overview:histogram}). Finally, if no qubit pair
can be mapped, the process alternates to map most referenced logical qubits to 
physical qubits with the best fidelity and lowest read error.


\section{Mapping}
\label{sec:mapping}

\subsection{Detecting Sub-Strings}

As first step, we extract a linearized trace from 
all the pairs of qubit indices appearing in 2-qubit
gates (e.g. $\langle 1, 2 \rangle$ from \texttt{CX 1,2}), and encoding
the pair as a $q_1 ~.~ Q_{device} + q_2$, where $Q_{device}$ is the number
of physical qubits in the quantum device, $q_1$ is the first component of the
pair, and $q_2$ the second one. The linearization allows us distinguish 
pairs $\langle i,j \rangle$ from  $\langle j,i \rangle$. We do note, however, that
all 2-qubit gate operations, regardless of their type, will be encoded in this
fashion.

Next, we proceed to determine whether there exists a subsequence
of QASM operations that repeats within the linearized two-qubit stream.
If a repeating subsequence does exist, we are interested in the longest,
non-overlapping one. In summary, we aim to find the {\em longest
repeating non-overlapping substring}. Then, the most relevant qubits within each
substring will be mapped first.

Eq.~\ref{eq:string-match} is used to compute the length of matching substrings
from a given string $S$, and storing the prefix lengths in a table $T$.  Table
$T$ is built from the bottom-up, computing first prefixes of length 1, then
prefixes of length 2 and so on.  It is easy to see that if $S_i \neq S_j$ then
the matching prefix length is zero; Otherwise the new prefix length effectively
increases by one.  Eq.~\ref{eq:string-match} can be easily implemented in an
iterative fashion; However, care must be taken to minimize the memory footprint
of $T$ in order to avoid incurring in a $O(n^2)$ space complexity, which for
QASM-based circuits soon consisting of millions of QOPS can result in
significant memory requirements.  To illustrate this point, a program with 100K
QASM operations, and encoding each qubit index pair as a 4-byte integer would
require $32.25GB$ while one with 500K quantum operations would require $931GB$.
Thus, when computing this table, we resort to keep only the two most recent
rows, keeping the maximum length of $T_{i,j}$, and storing the $i$ index at
each update to the maximum length.

\begin{eqnarray}
\label{eq:string-match}
T_{i,j} = 
\begin{cases}
T_{i-1,j-1} + 1 : & S_i = S_j \wedge j > i \wedge T_{i-1,j-1} < j - i \\
0            & otherwise
\end{cases}
\end{eqnarray}

\subsection{Global Frequencies}
\label{sec:globfreq}
Next, We compute for each qubit pair $\langle q_i, q_j \rangle$ the number of
times that such pair is used in two-qubit quantum gates. We denote this
value as $refs(\langle q_i, q_j \rangle)$. Then,
for each qubit $q_i$ we compute its total usage frequency as:

\begin{eqnarray*}
freq(q_i) = \sum_{q_j} refs(\langle q_i, q_j \rangle)
\end{eqnarray*}

In several instances, and
to prioritize the mapping of qubits.
we will also sort all neighboring qubits $q_j$,
for a given $q_i$, by their $refs$ value.

\subsection{Mapping Qubits from Sub-Strings}

\newcommand{\qset}[1]{$Q_{#1}$}
\newcommand{\qubit}[1]{$q_{#1}$}
\newcommand{\qlist}[1]{$L_{#1}$}
\newcommand{\pstr}[0]{$P_{str}$}
\newcommand{\lmap}{$M_{l2p}$}
\newcommand{\larr}{$\leftarrow$}
\newcommand{\myvar}[1]{$#1$}

\begin{algorithm}[!htb]
\caption{\label{algo:mapping-by-ss}Map Qubits by Sub-String (SS) Method}
\begin{algorithmic}[1]
\REQUIRE{\qset{phys}: Physical Qubit Set; \qset{log}: Logical Qubit Set; $P$ : QASM Program}
\ENSURE{\lmap{}: Logical-to-physical qubit layout map.}
\STATE /* Initialize pending list */
\STATE{\qset{pending} \larr{} Extract two-qubit indices from $P$;}
\STATE{\qset{avail} \larr{} \qset{phys}; }

\STATE{\myvar{edges} \larr{} Extract\_qubit\_pairs\_from\_program ($P$);}
\STATE{\pstr{} \larr{} linearize\_qubit\_pairs (\myvar{edges}); }
\WHILE{(not done)}
  \STATE{/* Build a histogram of edges (pairs of qubits) */}
  \STATE{$H$ \larr{} build\_qubit\_edge\_frequency (\myvar{edges}); }
  \STATE{/* Select High-Impact Qubits (hiq) from H */}
  \STATE{\qlist{hiq} \larr{} build\_high\_impact\_qubit\_list (\myvar{H});}
  \STATE{\myvar{str} \larr{} find\_lrnos (\pstr{});}
  \IF{length(\myvar{str}) $< 2$}
    \STATE{done \larr{} True;}
  \ENDIF
  \STATE{\myvar{ehist} \larr{} substring\_histogram (str);}
  \WHILE {(\qubit{phys} \larr{} get\_next\_physical\_qubit (\qset{avail}))}
    \IF{(\qubit{phys} is invalid)}
      \STATE{\textbf{break;}}
    \ENDIF
    \STATE{\qlist{phys} \larr{} $\{ q_{phys} \} \cup ( adj(q_{phys}) \cap Q_{avail}); $}
    \STATE{\qlist{log} \larr{} select\_logical\_qubits (\qset{pending}, \qlist{hiq}, \myvar{ehist});}
    \IF{size(\qlist{log}) < 2}
      \STATE{\textbf{break;}}
    \ENDIF
    \STATE{\lmap{} \larr map\_qubits (\qlist{log}, \qlist{phys});}
    \STATE{\qset{pending} \larr{} \qset{pending} - \qlist{log};}
    \STATE{\qset{avail} \larr{} \qset{avail} - \qlist{phys};}
    \STATE{sort \qset{avail} by number of available neighbors;}
  \ENDWHILE
  \IF{no qubits were mapped in previous loop}
    \STATE{\pstr{} \larr{} remove\_substring\_occurrences (\pstr{}, \myvar{str});}
    \STATE{sort \qset{avail} by number of available neighbors;}
    \STATE{\myvar{edges} \larr \myvar{edges} - $\{ \langle q_i,q_j \rangle : q_i ~ or ~ q_j \not\in Q_{pending} \}$; }
  \ENDIF
\ENDWHILE
\RETURN{\lmap{};}
\end{algorithmic}
\end{algorithm}

Algorithm \ref{algo:mapping-by-ss} presents the steps taken to map
a subset of logical qubits used in the input program to a set of physical
available qubits. The overall algorithm proceeds by finding
{\em longest repeating non-overlapping sub-strings} (LRNOS) in the quantum program,
and proceeding to map as many {\em high-impact qubits} (HIQ) as possible from it.
We define a HIQ as a qubit with a higher $freq$ than the mean, but restricted to the
current LRNOS.
To differentiate HIQ from other qubits, we compute the average number of times
that a logical qubit $q_i$ appears as the first operand of a two-qubit
gate operation (e.g. $CX~ q_i,~q_j$), as defined in Sec.\ref{sec:globfreq}. 
This number is then used as a minimum
threshold to properly identify any $HIQ$. In addition, we also require logical
qubits to be in the \qset{pending} set.

Qubit mapping proceeds at two levels, {\em rounds} and {\em steps}. Rounds 
allocate qubits for a given LRNOS, and are
controlled in the outermost while-loop of Alg.~\ref{algo:mapping-by-ss};
Steps perform finer-grain actions in the innermost while-loop, by first
selecting a leading physical qubit ($q_{phys}$) to be used as the {\em control} qubit to a
number of 2-qubit gate operations, followed by selecting the still available
physical qubits adjacent to the leading one ($adj(q_{phys})$).  A similar selection is done for
the logical qubits;
The lead logical qubit is selected from the \qlist{hiq}, after which
the $k = len(adj(q_{phys}))$ most often qubits $q_j$ such that $\langle q_i,q_j \rangle$
are chosen and added to $L_{log}$.
Once logical and physical
qubits have been chosen, we map each corresponding leading qubits and their
adjacent ones, assigning more heavily used logical qubits to the physical
qubits with the lowest error. The {\em steps} loop exits when either a leading
physical qubit cannot be found, or when the set of logical qubits, \qlist{log},
consists of only one qubit (the leading one). The latter scenario occurs
when non-HIQ qubits remain in the current LRNOS.
After exiting the {\em steps} loop, a few actions are taken to prepare for the
next LRNOS: sub-string occurrences of the current LRNOS are removed from \pstr{},
the set/list of available physical qubits (\qset{avail}) is sorted by the number
of available neighbors, while the list of linearized qubit pairs (\myvar{edges})
is updated by removing all edges that have been partially allocated either
because the source or target qubit are not pending anymore.


\subsection{Mapping Remaining 2-qubit gate Qubits}

\begin{algorithm}[!htb]
\caption{\label{algo:mapping-by-gf}Map Qubit by Global Frequency (GF)}
\begin{algorithmic}[1]
\REQUIRE{\qset{phys}: Physical Qubit Set; \qset{log}: Logical Qubit Set; $P$ : QASM Program}
\ENSURE{\lmap{}: Logical-to-physical qubit layout map.}
\STATE /* Initialize pending list */
\STATE{\qset{pending} \larr{} Extract two-qubit indices from $P$;}
\STATE{\qset{pending} \larr{} \qset{pending} - physical(\lmap{})}
\STATE{sort \qset{avail} by number of available physical neighbors;}
\STATE{\myvar{edges} \larr{} Extract\_qubit\_pairs\_from\_program ($P$);}
\WHILE{(\qset{pending} $\neq \varnothing$)}
  \STATE{\myvar{edges} \larr{} \myvar{edges} $\cap \{ \langle q_i, q_j \rangle : q_i, q_j \in Q_{pending} \}$ ;}
  \STATE{\myvar{ehist} \larr{} build\_histogram (\myvar{edges});}
  \IF{\myvar{ehist} is empty}
    \STATE{\textbf{break;}}
  \ENDIF
  \STATE{\qubit{phys} \larr{} get\_next\_physical\_qubit (\qset{avail});}
  \STATE{\qlist{phys} \larr{} $\{ q_{phys} \} \cup ( adj(q_{phys}) \cap Q_{avail}); $}
  \STATE{\qlist{log} \larr{} select\_logical\_qubits (\qset{pending}, \myvar{ehist});}
  \IF{size(\qlist{log}) < 2}
    \STATE{\textbf{break;}}
  \ENDIF
  \STATE{\lmap{} \larr map\_qubits (\qlist{log}, \qlist{phys});}
  \STATE{\qset{pending} \larr{} \qset{pending} - \qlist{log};}
  \STATE{\qset{avail} \larr{} \qset{avail} - \qlist{phys};}
\ENDWHILE
\RETURN{\lmap{};}
\end{algorithmic}
\end{algorithm}

\begin{table*}[!ht]
\begin{footnotesize}
\caption{\label{tab:qiskit:depth} Average of normalized depth variation by circuit size -- \qiskit{} compiler.}
\begin{tabular}{c|c|c|c|c|c|c||c|c|c|c|c|c}
\toprule
   & \multicolumn{6}{c||}{IBM Manhattan (65 qubit)} & \multicolumn{6}{c}{IBM Kolkata (27 qubit)} \\
C1 & C2 & C3 & C4 & C5 & C6 & C7 & C8 & C9 & C10 & C11 & C12 & C13 \\
Group [Size Range] : & SS$_b$ & SS$_o$ & GF$_b$ & GF$_o$ & GSF$_b$ & GSF$_o$ & SS$_b$ & SS$_o$ & GF$_b$ & GF$_o$ & GSF$_b$ & GSF$_o$ \\
(Instances):       &  &  &  &  &  &  &  &  &  &  &  &  \\
\toprule
 G1  [ 1:   1K] (105) &  1.01 &  1.06 &  1.07 &  1.01 &  0.99 &  1.02  &  1.03 &  1.03 &  1.07 &  1.08 &  0.95 &  1.08\\
 G2 [ 1K:   5K] (15) &  1.11 &  1.03 &  1.10 &  1.02 &  1.12 &  1.03   &  1.11 &  1.02 &  1.11 &  1.00 &  1.12 &  1.02\\
 G3 [ 5K:  10K] (6) &  1.13 &  0.98 &  1.15 &  0.98 &  1.14 &  0.98    &  1.15 &  0.99 &  1.13 &  0.99 &  1.14 &  1.00\\
 G4 [10K:  20K] (7) &  1.13 &  1.00 &  1.13 &  1.02 &  1.12 &  1.02    &  1.13 &  1.01 &  1.12 &  1.02 &  1.12 &  1.01\\
 G5 [20K:  30K] (5) &  1.13 &  0.95 &  1.12 &  0.97 &  1.13 &  0.97    &  1.14 &  1.00 &  1.16 &  1.01 &  1.17 &  1.00\\
 G6 [30K:  40K] (5) &  1.13 &  1.00 &  1.11 &  0.99 &  1.13 &  1.00    &  1.13 &  0.99 &  1.12 &  1.00 &  1.13 &  1.02\\
 G7 [40K: 100K] (5) &  1.13 &  0.98 &  1.13 &  0.98 &  1.14 &  1.00    &  1.12 &  0.97 &  1.14 &  1.01 &  1.11 &  0.99\\
 G8 [100K: 250K] (7) &  1.10 &  1.00 &  1.11 &  0.98 &  1.10 &  1.00    &  1.11 &  0.99 &  1.11 &  1.00 &  1.12 &  1.00\\
 G9 [250K: 1M] (2) &  1.09 &  0.98 &  1.09 &  0.98 &  1.09 &  0.98    &  1.16 &  1.05 &  1.14 &  1.05 &  1.16 &  1.05\\
\bottomrule
\end{tabular}
\end{footnotesize}
\end{table*}

\begin{table*}[!htb]
\begin{footnotesize}
\caption{\label{tab:tket:depth} Average of normalized depth variation by circuit size -- \tket{} compiler.}
\begin{tabular}{c|c|c|c|c|c|c||c|c|c|c|c|c}
\toprule
   & \multicolumn{6}{c||}{IBM Manhattan (65 qubit)} & \multicolumn{6}{c}{IBM Kolkata (27 qubit)} \\
C1 & C2 & C3 & C4 & C5 & C6 & C7 & C8 & C9 & C10 & C11 & C12 & C13 \\
Group [Size Range] & SS$_b$ & SS$_o$ & GF$_b$ & GF$_o$ & GSF$_b$ & GSF$_o$ & SS$_b$ & SS$_o$ & GF$_b$ & GF$_o$ & GSF$_b$ & GSF$_o$ \\
(Instances):       &  &  &  &  &  &  &  &  &  &  &  &  \\
\toprule
G1 [ 1:   1K] (105) &  0.66 &  0.78 &  0.71 &  0.88 &  0.64 &  0.75  &  0.68 &  0.80 &  0.70 &  0.89 &  0.63 &  0.72\\
G2 [ 1K:  5K] (15)  &  0.89 &  0.97 &  0.89 &  1.02 &  0.88 &  0.96  &  0.79 &  0.89 &  0.81 &  0.90 &  0.80 &  0.89\\
G3 [ 5K:  10K] (6)  &  0.85 &  0.97 &  0.87 &  0.98 &  0.84 &  1.03  &  0.89 &  0.97 &  0.86 &  1.03 &  0.88 &  1.05\\
G4 [ 10K: 20K] (7)  &  0.93 &  1.03 &  0.91 &  1.05 &  0.97 &  1.10  &  0.94 &  1.04 &  0.90 &  1.03 &  0.96 &  1.05\\
G5 [ 20K: 30K] (5)  &  0.99 &  1.10 &  1.01 &  1.12 &  0.97 &  1.06  &  0.97 &  1.06 &  0.97 &  1.07 &  0.98 &  1.08\\
G6 [ 30K: 40K] (5)  &  0.93 &  1.04 &  0.94 &  1.09 &  0.93 &  1.08  &  0.92 &  1.00 &  0.94 &  1.05 &  0.94 &  1.02\\
G7 [ 40K: 100K] (5) &  0.92 &  1.02 &  0.92 &  1.03 &  0.92 &  1.02  &  0.86 &  0.96 &  0.88 &  0.98 &  0.85 &  0.96\\
G8 [100K: 250K] (7) &  0.92 &  1.05 &  0.93 &  1.04 &  0.93 &  1.04  &  0.93 &  1.04 &  0.92 &  1.03 &  0.94 &  1.05\\
G9 [250K: 1M] (2)   &  0.92 &  1.01 &  0.98 &  1.09 &  0.92 &  1.01  &  0.86 &  0.97 &  0.87 &  0.96 &  0.86 &  0.97\\
\bottomrule
\end{tabular}
\end{footnotesize}
\end{table*}

Next, we present a second algorithm, inspired in MUQUT's \cite{muqut} k-vertex
topology graph extraction, and which complements the sub-string based qubit
allocation method introduced in the previous section. 
Algorithm \ref{algo:mapping-by-gf} is a more direct version of 
Algorithm \ref{algo:mapping-by-ss}. It only attempts to map qubits
by considering the histogram (global frequency) of logical qubits used
in 2-qubit gates. Consequently, the assignment of qubits is performed
in a single loop, choosing first a leading physical qubit (and indirectly,
its adjacent physical ones), followed by selecting the leading logical qubit 
together with its neighbors. The actual mapping proceeds in the same fashion
as in the first algorithm. The loop stops whenever we find that the updated
global histogram \myvar{ehist} has no pending edges (NOTE: qubits can still
be pending, but they will no form a pair $\langle q_i, q_j \rangle$
where both qubits are in the \qset{pending}).
At the end of each iteration, both the logical pending qubit set (\qset{pending})
and the physical available qubit set (\qset{avail}) are accordingly updated.

\noindent
\begin{table}[h]
\begin{footnotesize}
\caption{\label{tab:testbed}Experimental Testbed}.
\noindent
\begin{tabular}{l|l|l}
\toprule
\multicolumn{3}{c}{Compilers} \\
\toprule
                & Base Flags / Passes   & Opt Flags / Passes \\ 
\midrule
\tket{}   &  identity placement   +         &  Same as base +     \\
v0.13.0     &  routing +                   &  FullPeepholeOptimize + \\
            &  IBM rebase +                   &   NoiseAwarePlacement  \\
            &  validity predicate             &  (Only for Original QASM) \\ 
\midrule
\qiskit{}  & layout\_method=                &  layout\_method='sabre' +  \\
             & ~~~~~~~~~~~'trivial' +          &  routing\_method='sabre' +  \\
                &    opt-level=2                 &  scheduling\_method=None +  \\ 
 v0.23.6            &                                &  translation\_method=       \\
             &                                &   ~~~~~~~~~~ 'translator' + \\
             &                                &  opt-level=2   \\
\bottomrule
\end{tabular}
\end{footnotesize}
\end{table}

\subsection{Mapping Remaining 1-qubit gate Qubits}
If after mapping qubits with Algorithm \ref{algo:mapping-by-ss} -- \ref{algo:mapping-by-gf}
no unmapped pair $\langle q_i, q_j \rangle$ can be found, we then proceed
to map qubits based solely on their single-qubit gate error and read error rates, which
are obtained from the physical parameter files associated to each
quantum device.

\section{Evaluation}
\label{sec:evaluation}

We implemented the qubit mapping methods described in Sec.\ref{sec:mapping}
as a single standalone tool which takes quantum circuits in OpenQASM and emits
a remapped program also in OpenQASM. The compilers used, together with their
version and options are listed in Table \ref{tab:testbed}. 
Devices used are the IBM Manhattan (65-qubit, 72 edges) and Kolkata (27-qubit, 28 edges) 
\cite{ibm-services}.  Physical parameter files were retrieved on August 22,
2021.

\paragraph{Evaluated Circuits:}
We use the \texttt{IBM-QX QASM} benchmark set \cite{jku-ibmqx-web},
which has been extensively used
in prior works \cite{jku.date.2018,sabre,tket,olsq}.
All circuits in this collection utilize up to 16 qubits.
The suite consist of 158 quantum circuits, of which we use all but the
{\em ground-state-estimation} circuit owing to compilers removing
all but a handful of operations from a program with $390K$ quantum operations.
The average circuit size (in number of quantum gates) of the remaining
157 circuits is $18,714.5$ operations, with a standard deviation of $62,863$.
However, 105 of the circuits present 1000 or fewer operations. We thus organize
our results in 9 groups. The specific ranges of circuit sizes can be found
on the leftmost column of Table \ref{tab:qiskit:depth}--\ref{tab:tket:volume}.
The new mappings were collected in
a 32-core AMD Thread-Ripper server, with 128 GB DRAM, and operating with Ubuntu 18.
The total time to generate all 157 layouts, for the Manhattan topology, with our
SS, GF and GSF mappings was 79.2sec, 79.4sec and 159.3sec, respectively.

\begin{table*}[!ht]
\begin{footnotesize}
\caption{\label{tab:qiskit:volume} Average of normalized gates variation by circuit size: \qiskit{} compiler.}
\begin{tabular}{c|c|c|c|c|c|c||c|c|c|c|c|c}
\toprule
   & \multicolumn{6}{c||}{IBM Manhattan (65 qubit)} & \multicolumn{6}{c}{IBM Kolkata (27 qubit)} \\
C1 & C2 & C3 & C4 & C5 & C6 & C7 & C8 & C9 & C10 & C11 & C12 & C13 \\
Group [Size Range] & SS$_b$ & SS$_o$ & GF$_b$ & GF$_o$ & GSF$_b$ & GSF$_o$ & SS$_b$ & SS$_o$ & GF$_b$ & GF$_o$ & GSF$_b$ & GSF$_o$ \\
(Instances):       &  &  &  &  &  &  &  &  &  &  &  &  \\
\toprule
G1 [ 1:   1K] (105) &  0.55 &  1.04 &  0.63 &  0.98 &  0.53 &  0.98  &  0.57 &  1.01 &  0.63 &  1.00 &  0.49 &  1.00\\
G2 [ 1K:  5K] (15)  &  0.63 &  1.04 &  0.63 &  1.04 &  0.64 &  1.05   &  0.64 &  1.00 &  0.64 &  0.99 &  0.64 &  1.00\\
G3 [ 5K:  10K] (6)  &  0.63 &  0.98 &  0.65 &  0.97 &  0.64 &  0.97    &  0.65 &  0.98 &  0.62 &  0.99 &  0.64 &  1.00\\
G4 [ 10K: 20K] (7)  &  0.64 &  1.00 &  0.64 &  1.00 &  0.63 &  1.02    &  0.64 &  1.01 &  0.63 &  1.02 &  0.64 &  1.01\\
G5 [ 20K: 30K] (5)  &  0.64 &  0.90 &  0.62 &  0.94 &  0.64 &  0.93    &  0.64 &  0.99 &  0.66 &  1.01 &  0.68 &  1.01\\
G6 [ 30K: 40K] (5)  &  0.64 &  1.00 &  0.63 &  1.01 &  0.64 &  1.00    &  0.63 &  1.00 &  0.63 &  1.00 &  0.63 &  1.04\\
G7 [ 40K: 100K] (5) &  0.64 &  0.98 &  0.64 &  0.98 &  0.66 &  0.99    &  0.62 &  0.92 &  0.65 &  1.01 &  0.62 &  0.97\\
G8 [100K: 250K] (7) &  0.65 &  1.01 &  0.66 &  1.00 &  0.65 &  1.01    &  0.64 &  0.98 &  0.64 &  0.99 &  0.66 &  1.01\\
G9 [250K: 1M] (2)   &  0.60 &  0.92 &  0.60 &  0.91 &  0.60 &  0.92    &  0.65 &  1.02 &  0.63 &  0.98 &  0.65 &  1.02\\
\bottomrule
\end{tabular}
\end{footnotesize}
\end{table*}

\begin{table*}[!htb]
\begin{footnotesize}
\caption{\label{tab:tket:volume} Average of normalized gates variation by circuit size -- \tket{} compiler.}
\begin{tabular}{c|c|c|c|c|c|c||c|c|c|c|c|c}
\toprule
   & \multicolumn{6}{c||}{IBM Manhattan (65 qubit)} & \multicolumn{6}{c}{IBM Kolkata (27 qubit)} \\
C1 & C2 & C3 & C4 & C5 & C6 & C7 & C8 & C9 & C10 & C11 & C12 & C13 \\
Group [Size Range] & SS$_b$ & SS$_o$ & GF$_b$ & GF$_o$ & GSF$_b$ & GSF$_o$ & SS$_b$ & SS$_o$ & GF$_b$ & GF$_o$ & GSF$_b$ & GSF$_o$ \\
(Instances):       &  &  &  &  &  &  &  &  &  &  &  &  \\
\toprule
G0 [ 1:   1K] (105) &  0.52 &  0.64 &  0.59 &  0.87 &  0.48 &  0.64  &  0.52 &  0.69 &  0.56 &  0.81 &  0.44 &  0.57\\
G1 [ 1K:  5K] (15)  &  0.78 &  0.91 &  0.79 &  0.97 &  0.78 &  0.91  &  0.71 &  0.82 &  0.71 &  0.83 &  0.71 &  0.82\\
G2 [ 5K:  10K] (6)  &  0.78 &  0.96 &  0.79 &  0.96 &  0.76 &  1.02  &  0.80 &  0.99 &  0.79 &  1.03 &  0.80 &  1.10\\
G3 [ 10K: 20K] (7)  &  0.89 &  1.04 &  0.88 &  1.06 &  0.91 &  1.11  &  0.88 &  1.04 &  0.86 &  1.03 &  0.91 &  1.05\\
G4 [ 20K: 30K] (5)  &  0.94 &  1.11 &  0.96 &  1.12 &  0.94 &  1.08  &  0.92 &  1.06 &  0.91 &  1.07 &  0.93 &  1.07\\
G5 [ 30K: 40K] (5)  &  0.88 &  1.07 &  0.89 &  1.10 &  0.88 &  1.09  &  0.88 &  1.01 &  0.89 &  1.05 &  0.89 &  1.03\\
G6 [ 40K: 100K] (5) &  0.85 &  1.01 &  0.85 &  1.02 &  0.87 &  1.00  &  0.81 &  0.93 &  0.80 &  0.95 &  0.78 &  0.94\\
G7 [100K: 250K] (7) &  0.88 &  1.06 &  0.88 &  1.06 &  0.88 &  1.05  &  0.87 &  1.04 &  0.85 &  1.02 &  0.88 &  1.05\\
G8 [250K: 1M] (2)   &  0.86 &  1.02 &  0.93 &  1.11 &  0.86 &  1.02  &  0.80 &  0.95 &  0.79 &  0.93 &  0.80 &  0.95\\
\bottomrule
\end{tabular}
\end{footnotesize}
\end{table*}

\subsection{General Trends}
\label{sec:eval:summary}
We focus our analysis on the {\em circuit depth} (number of gates in the program's
critical path) and on the {\em gate volume} metric
(number of gates in the program). However, as in some cases the impact on these
metrics can be small relative to the original circuit size, we take the approach
of previous works \cite{sabre,olsq,tket} and consider only the metric variation,
i.e., the difference between the metric obtained after compilation and the
original metric value (circuit depth or gate volume). 

We show the summary of our results on Tables \ref{tab:qiskit:depth} --
\ref{tab:tket:volume}. The first column in all four tables indicates the group number
(range of circuit sizes prior to compilation) and the number of circuit instances
meeting the criterion. The next 12 columns list the average of our normalized results,
i.e. the ratio of the metric obtained with the baseline configuration by 
the metric value obtained when using our mapping (or a combination of them).
Column names denote the combination of our 3 mappings (SS, GF, GSF) in tandem
with two different compiler configurations, $b$ for {\em base} and $o$ for {\em optimized}.
As previously discussed, GSF is the back-to-back application of our two mapping strategies, 
Sub-String (SS)
followed by Global Frequencies (GF). Columns C2--C7 correspond to results collected
on IBM Manhattan while columns C8--C13 correspond to IBM Kolkata.

\paragraph{Impact on Circuit Depth}
From Table \ref{tab:qiskit:depth} and Table \ref{tab:tket:depth} we observe
that our qubit mappings are especially beneficial when used in
conjunction with the \qiskit{} compiler, while benefits when utilizing \tket{}
are more noticeable for groups G3--G9, the larger circuit classes. We recall
that average values above 1.00 represent an improvement on the metrics. 
We can also observe an interesting phenomenon between groups G1 and G2 on Table
\ref{tab:qiskit:depth}, which is an inversion in trends; On group G1
configurations \msso{}, \mgfo{} and \mgsfo{} outperform 
\mssb{}, \mgfb{} and \mgsfb{}. Then, starting with group G2, the {\em base}
configurations always outperform the {\em optimized} ones.
This demonstrates that, \qiskit{} is currently better tuned for smaller circuits.
In contrast, the same effect is not detected in Table \ref{tab:tket:depth},
where the {\em optimized} configurations always outperform the {\em base} ones.
Nonetheless, while a similar inversion does not manifest with \tket{},
first signs of improvements appear on group G3, and become more consistent
on groups G4 -- G9. This suggests that \tket{}'s qubit placement method
is better tuned for circuits with up to 10K gates.

Next, we note that our results with \qiskit{} are somewhat better on the
Kolkata topology, where we achieve an average circuit depth improvement 
of up to $1.17\times$  on G5 and $1.16\times$ on G9 with \mgsfo{}. These are good
examples of when \mss{} and \mgf{} can work in tandem to obtain better overall
results. In comparison, results on Manhattan exhibit less variation among groups,
and averaging $1.13\times$ improvement in several groups.
Notable improvements with \tket{} on Manhattan can be seen on group G4,
where \mgsfo{} ($1.10\times$) clearly outmatches \msso{} ($1.03\times$) and \mgfo{}
($1.05\times$). Finally, we note a modest decline on the effectiveness of our techniques
for groups G7--G9 on Manhattan, and a more pronounced one on Kolkata for the
same groups.


\begin{figure*}[!htb]
\begin{subfigure}{0.48\linewidth}
\includegraphics[angle=0,width=\linewidth]{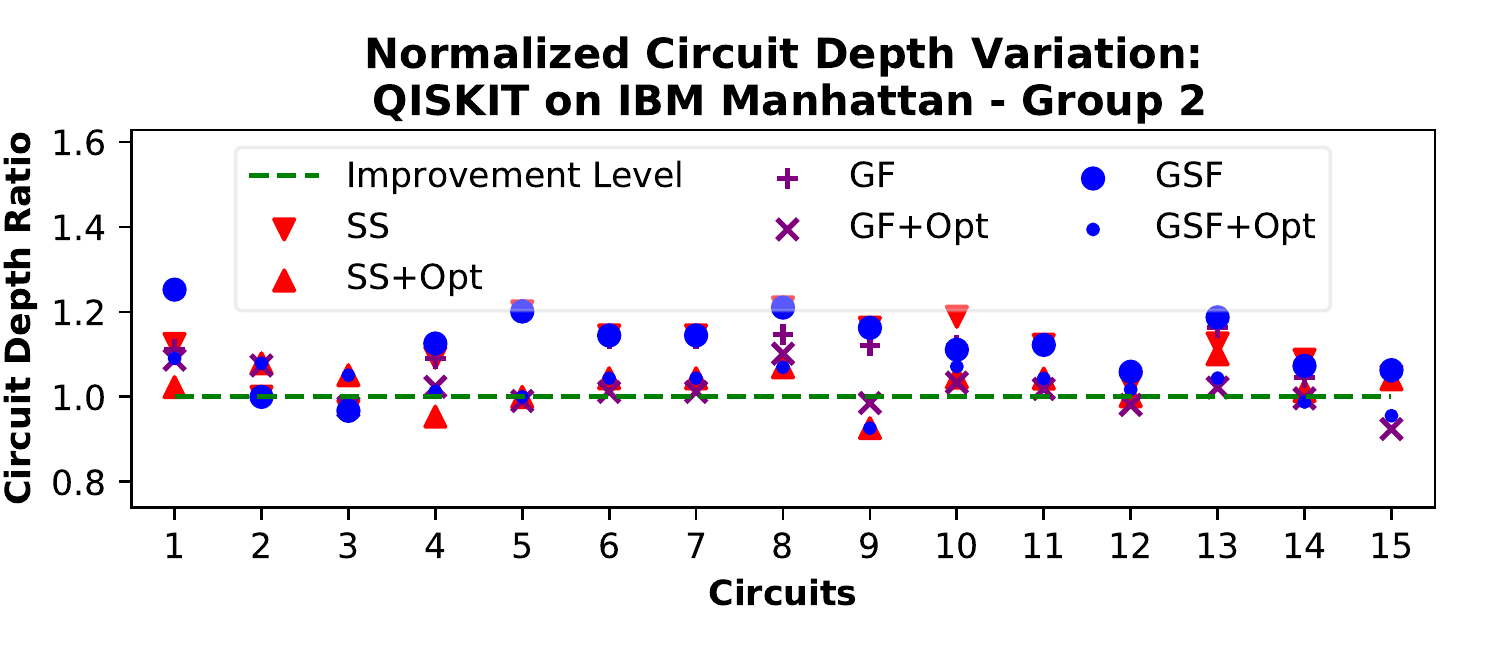}
\caption{\label{fig:qiskit:manhattan:g2} Circuit depth: \qiskit{}, Manhattan, group G2.}
\end{subfigure}
\begin{subfigure}{0.48\linewidth}
\includegraphics[angle=0,width=\linewidth]{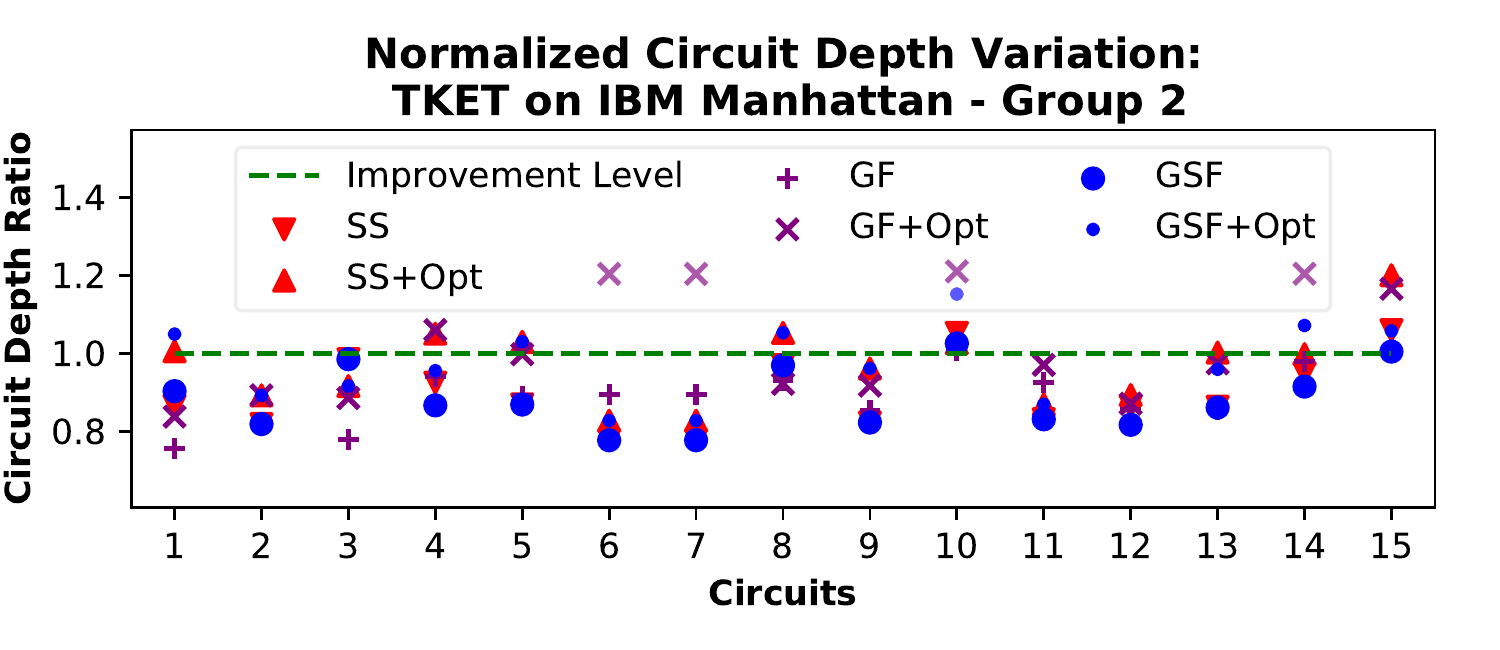}
\caption{\label{fig:tket:manhattan:g2} Circuit depth: \tket{}, Manhattan, group G2.}
\end{subfigure}
\begin{subfigure}{0.48\linewidth}
\includegraphics[angle=0,width=\linewidth]{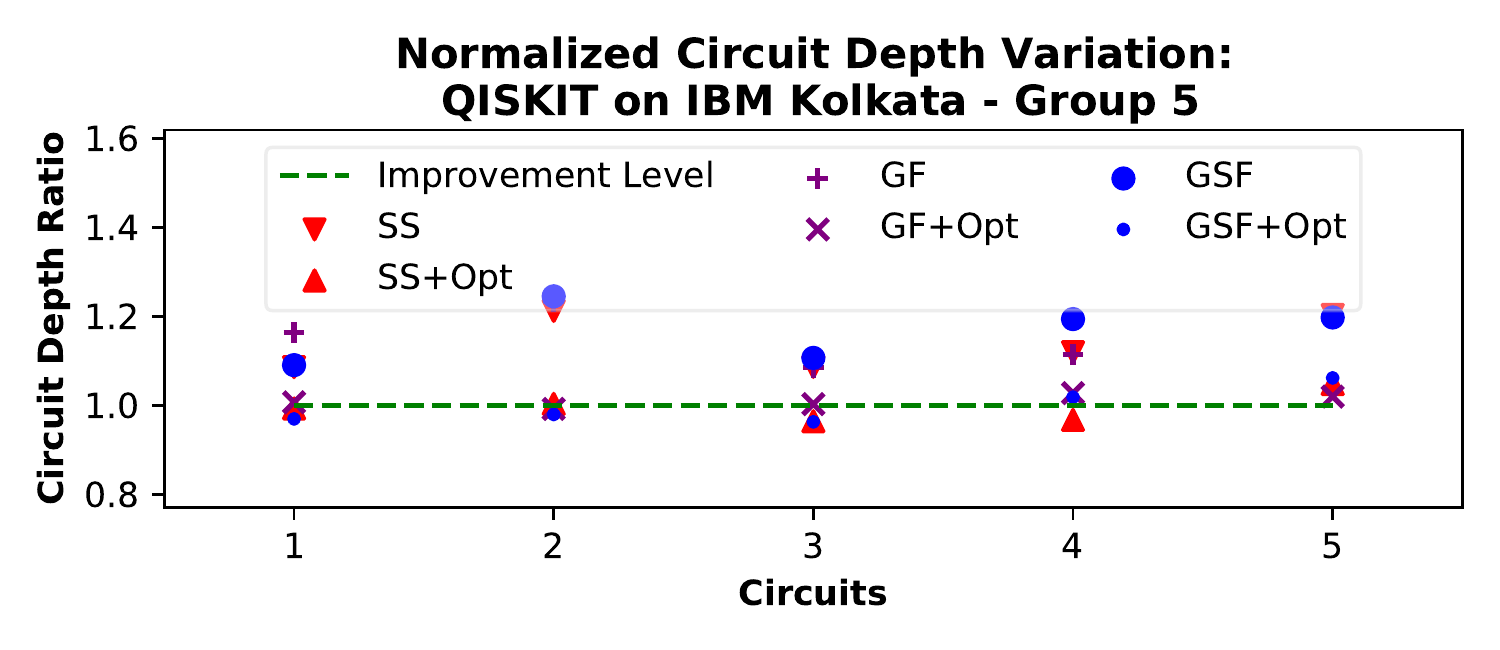}
\caption{\label{fig:qiskit:kolkata:g5} Circuit depth: \qiskit{}, Kolkata, group G5.}
\end{subfigure}
\begin{subfigure}{0.48\linewidth}
\includegraphics[angle=0,width=\linewidth]{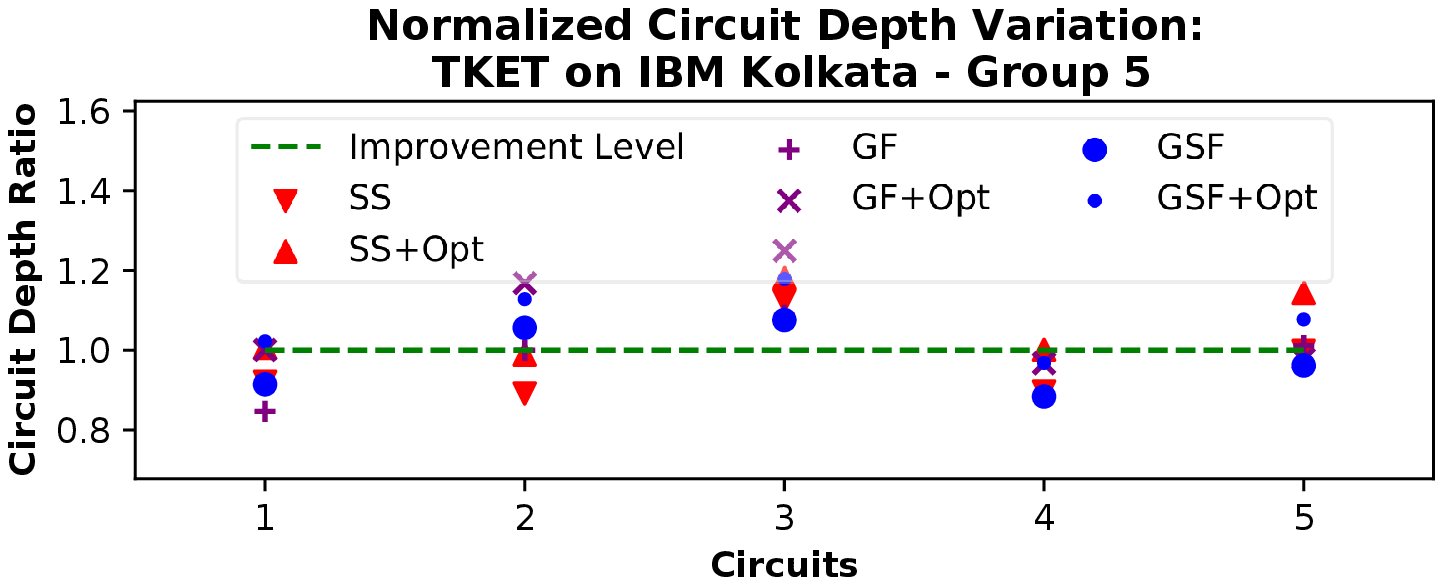}
\caption{\label{fig:tket:kolkata:g5} Circuit depth: \tket{}, Kolkata, group G5.}
\end{subfigure}
\caption{\label{fig:groups:depth}Normalized circuit depth variation in IBM-QX, groups 2 and 5.}
\end{figure*}

\begin{figure*}[!htb]
\begin{subfigure}{0.48\linewidth}
\includegraphics[angle=0,width=\linewidth]{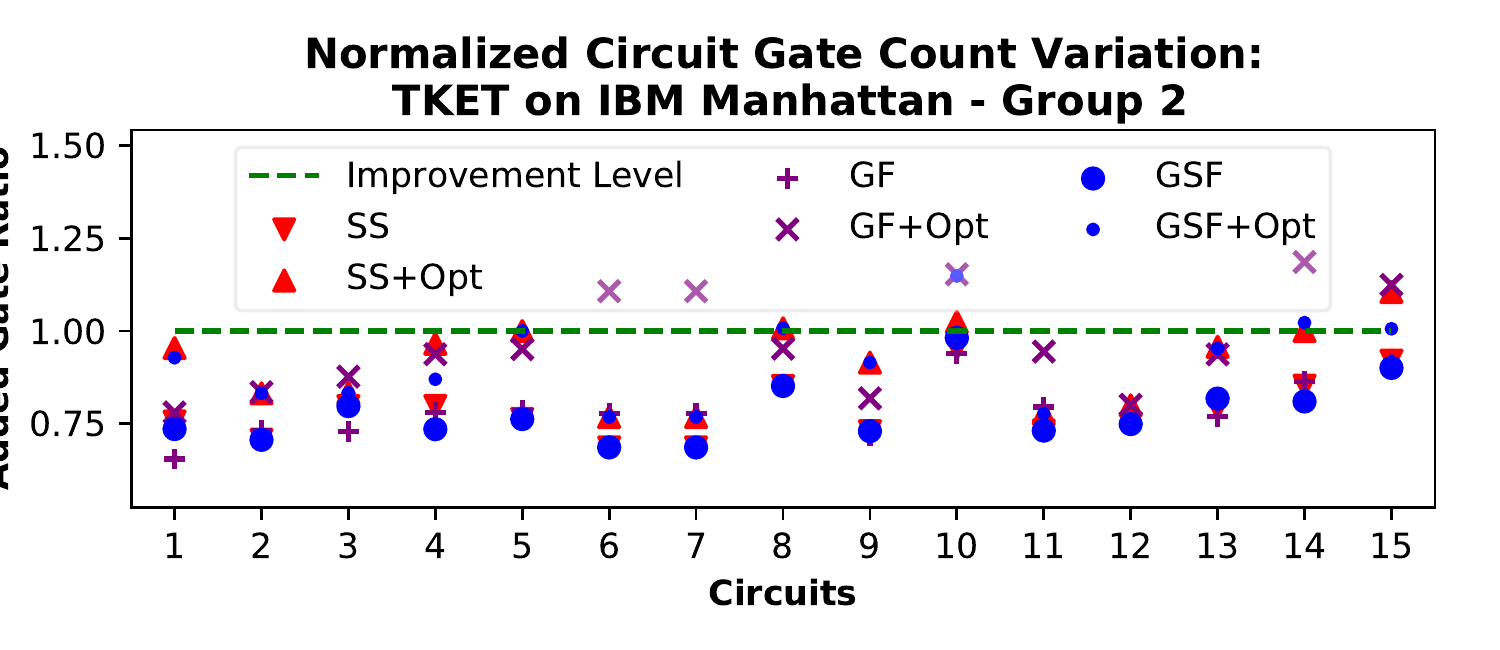}
\caption{\label{fig:tket:volume:manhattan:g2} Gate volume: \tket{}, Manhattan, group G2.}
\end{subfigure}
\begin{subfigure}{0.48\linewidth}
\includegraphics[angle=0,width=\linewidth]{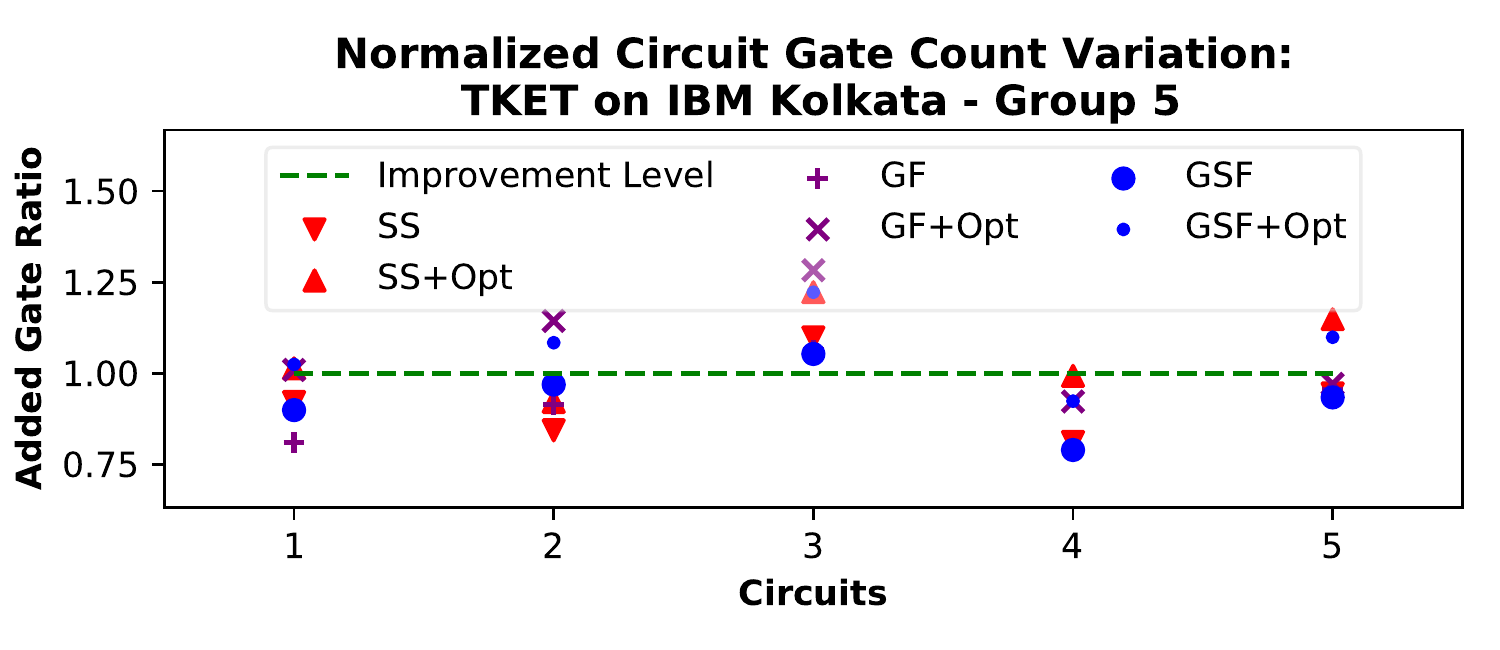}
\caption{\label{fig:tket:volume:kolkata:g5} Gate volume: \tket{}, Kolkata, group G5.}
\end{subfigure}
\caption{\label{fig:groups:volume}Normalized gate volume variation in IBM-QX, groups 2 and 5, only \tket{} compiler.}
\end{figure*}

\paragraph{Impact on Gate Volume}
Tables \ref{tab:qiskit:volume}--\ref{tab:tket:volume} summarize our
results on the gate count variation using both compilers and topologies.
The first observation we make is related to the effect of the compiler
configurations used {\em base} and {\em optimized}; Our methods were
particularly impactful on the circuit depth when using the \qiskit{}
compiler. Sadly, this appears to come with a penalty that manifests on the 
gate volume, where the {\em base} configurations are consistently outperformed
by the {\em optimized} ones. The phenomenon can be interpreted as additional
gates being introduced out of the critical path, but often enough to
yield a 30\%--40\% gap between the {\em base} and {\em optimized} configurations.
This trend is not observed when using the \tket{} compiler, where 
improvements on both the circuit depth and gate volume correlate.
Overall, we also remark that improvements in the gate volume when
using \qiskit{} only allow to match the baseline, whereas with \tket{}
improvements are of the same order as on the circuit depth.
For instance, on Manhattan's group G4, the aggregate improvement
when using \mgsfo{} (i.e. \msso{} + \mgfo{}) is of the order of $1.11\times$,
besting the $1.04\times$ of \msso{} and $.106\times$ of \mgfo{}.
In a similar fashion, \tket{} on Kolkata's group G3 achieves an average $1.10\times$
improvement in the gate volume, exceeding the $0.99\times$ of \msso{} and
$1.03\times$ of \mgfo{}.

\begin{figure*}[!htb]
\includegraphics[angle=0,width=\linewidth]{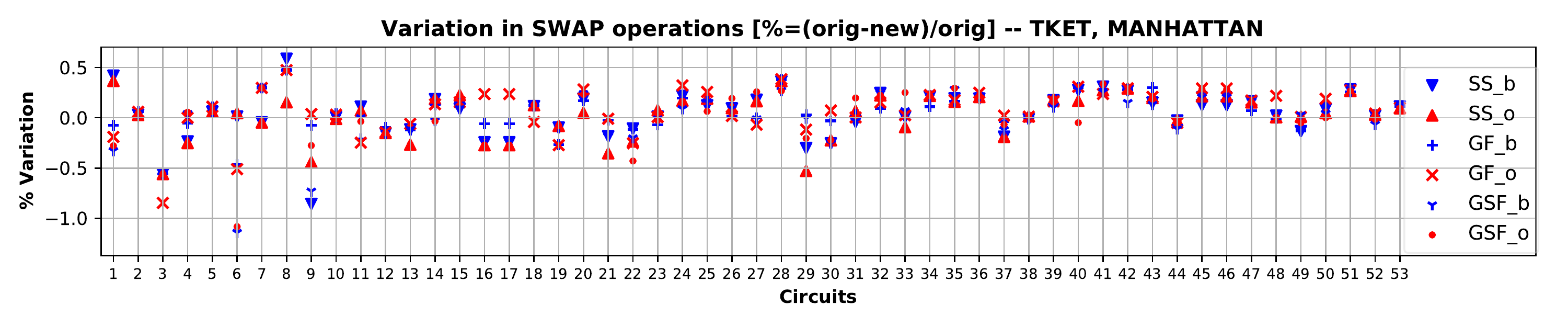}
\includegraphics[angle=0,width=\linewidth]{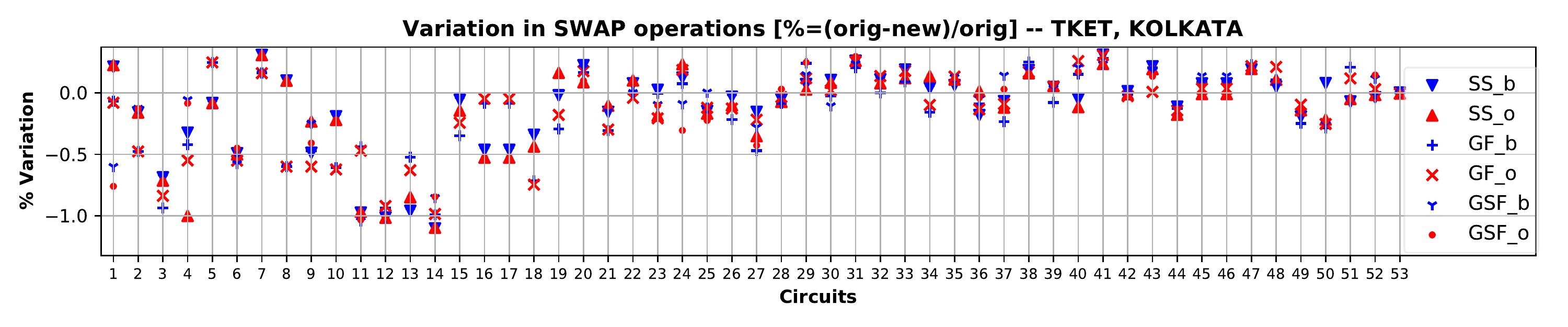}
\caption{\label{fig:swap}Impact of new qubit mappings on SWAP insertion with \tket{}: Manhattan (top) and Kolkata (bottom) devices.}
\end{figure*}

\subsection{Zooming into individual groups}
\label{sec:eval:individual}
We now turn our attention to understand the impact of our qubit mapping methods
within each circuit size group. Due to the large number of groups and
configurations, we focus on two groups, G2 and G5. Figure
\ref{fig:groups:depth} displays the normalized circuit depth for the 4
corresponding configurations of topologies and compilers.  We can see the
individual behavior of the 15 circuits comprising group G2.  Our mappings used with
the \qiskit{} compiler clearly provide better benefits than when using \tket{}
on Manhattan's group G2 (Figure \ref{fig:qiskit:manhattan:g2}). 
More specifically, we can notice that \mgsfb{}
constantly achieves the best improvement, with only one exception, \mssb{} at
x-tick 10. We also observe that \mgsfb{} outperforms \mgsfo{} in all but two
situations (x-tick 2 and 3). Results for the same topology and group, but this
time using \tket{} are not as good (Figure \ref{fig:tket:manhattan:g2}); 
The best improvements are obtained by the \mgfo{} mapping in 4 cases, while the
vast majority of data points fall below the improvement level. We believe this
is owed to \tket{} being more highly tuned for smaller circuits, i.e. groups G1
-- G3.  Very similar trends can be observed for group 5 on Kolkata, with
\qiskit{} achieving close to $1.2\times$ improvement on the circuit depth in
several instances (Figure \ref{fig:qiskit:kolkata:g5}), and producing benefits for 3 of
the 5 circuits with \tket{} (Figure \ref{fig:tket:kolkata:g5}).
Next, we shift our focus to analyze the impact of our mapping techniques on the
gate volume. These results can be seen in Fig.\ref{fig:groups:volume}.  In
general, we see that the {\em optimized} circuits practically always obtain
better gate volumes for
both topologies and both selected groups (G2 and G5). 
Interestingly, we also note that \tket{}'s gate 
volume results are generally worse on smaller circuits than with larger ones.
This observation aligns well with results on the circuit depth metric previously discussed.
Lastly, the 1.63\% reported improvement (Sec.~\ref{sec:intro}) in gate volume
was obtained in group G3, Kolkata device, with \tket{}.


\subsection{Impact on SWAP operations}
\label{sec:eval:swap-impact}
To conclude our evaluation, we present in Fig.\ref{fig:swap} plots showing
the impact of our mapping strategies over quantum circuits consisting
from 500 QOPS to 100K QOPS. 
This range spans parts of group G1 and all of
groups G2 -- G7. 
Data points are sorted by their original circuit size, i.e. the number of QOPS
prior to compilation.
Data clearly demonstrates that our qubit mapping approach
significantly reduces the number of SWAPs when used with \tket{}, nearing 
50\% SWAP reduction in a few occasions, and consistently reaching 30\% to 40\%
reduction for large enough circuits. Especially noticeable is the trend
change after the first 10 circuits,
where our results experience a marked improvement on Kolkata, and especially on Manhattan,
as the circuit size increases.
In addition, the fact that all our mapping variants improve collectively 
further strengthens the premise that \tket{} is currently better suited for small circuits
and for topologies with lower qubit counts. In contrast, improvements with
\qiskit{} are more modest, reaching up to 15\% improvement on Manhattan and 16\%
reduction on Kolkata (NOTE: Plot omitted due to space limitations).
We also note that the {\em optimized} configurations (the ones with suffix $_o$)
consistently achieve better results than their {\em base} counterparts.



\section{Related Work}
\label{sec:related}

The problem of qubit allocation has gained strong interest
in the last few years, largely on account of its impact on
on several quality metrics that affect the performance and
behavior of a quantum program \cite{murali.asplos.2019}. However, this problem
has been typically studied in conjunction with the {\em layout synthesis} task
(also known as {\em layout routing}),
which repairs a quantum program by moving qubit states onto physically
adjacent qubits to meet topological constraints \cite{maslov.tcad.2008}. Furthermore,
this problem has been proven to be NP-complete \cite{olsq,siraichi.cgo.2018}.
As a result, exhaustive enumeration and search techniques such as JKU's \cite{jku.date.2018},
while efficient on small circuits and small devices, will ultimately
not scale with programs expected to have millions of quantum gates.
Heuristics such as SABRE \cite{sabre} and WPM \cite{siraichi.cgo.2018}, 
which perform multiple 
passes on the input program, offer better scalability prospects.
In addition to SABRE, WPM, JKU, TKET line placement \cite{tket}, BMT \cite{siraichi.oopsla.2019},
Google's Cirq \cite{cirq.zenodo} greedy routing
and IBM's {\em Dense Layout} and {\em Stochastic Swap} \cite{qiskit},
also perform qubit placement by traversing the DAG quantum graph, 
and performing local decisions on network layers or qubit paths.
As an alternative, several works have also attempted to tackle this problem
using Integer Linear Programming and Boolean Satisfiability Solvers 
\cite{murali.asplos.2019,olsq,wille.dac.2019,murali.asplos.2020,muqut}.
These approaches have been successful in finding high-quality solutions 
to improve program success rates with noise-aware mappings.
However, they still attempt to map individual gate instances, often in combination
with gate scheduling. Consequently, they are more apt
for small circuits and small devices.
In contrast, the methods proposed in this work, in a spirit similar to \cite{wqcs21},
focus exclusively on the logical-to-physical
assignment of qubits, deferring the SWAP insertion job to later passes.

\section{Conclusion and Future Directions}
\label{sec:conclusion}

We have presented two qubit mapping techniques, which can be used
independently and together. We assume a decoupled strategy for qubit mapping,
where pairs of offending qubits not physically connected will be
repaired by a later compiler pass. More, our approach, based on detecting
repetitive string patterns, often combined with global frequency detection,
offers strong benefits with the \qiskit{} and \tket{} compilers, showing
improvement on quality metrics such as circuit depth, gate volume and 
SWAP count. As future work, we plan to further exploit string-oriented
quantum optimizations to find patterns that allow for the 
decomposition of large quantum programs, and leveraging repetitive
patterns to better exploit locality.


\newcommand{\xsederesource}[1]{Stampede2}
\newcommand{\xsedesp}[1]{Texas Advanced Computer Center (TACC)}
\newcommand{\xsedeallocid}[1]{CCR190043}


\bibliographystyle{ACM-Reference-Format}
\bibliography{allrefs}

\end{document}